\documentclass[12pt]{article}
\usepackage[english]{babel}
\usepackage[utf8]{inputenc}
\usepackage[T1]{fontenc}
\usepackage{authblk}
\usepackage{graphicx}
\usepackage{amsmath,mathrsfs,dsfont}
\usepackage{amsthm}
\usepackage[all]{xy}

\usepackage{graphicx,subfigure}
\usepackage{tensor}
\usepackage{verbatim}
\usepackage{extarrows}
\usepackage{nicefrac}
\usepackage{amsfonts}
\usepackage{amssymb,latexsym}
\usepackage{color}
\usepackage{mathrsfs,dsfont}
\usepackage{appendix}
\usepackage{slashed}
\usepackage{cite}
\usepackage{comment}

\newcommand{\be}{\begin{equation}}
\newcommand{\ee}{\end{equation}}
\newcommand{\beqa}{\begin{eqnarray}}
\newcommand{\eeqa}{\end{eqnarray}}

\definecolor{verde}{cmyk}{.83,.21,1,.08}

\begin{document}

\centerline{\LARGE  Dark Energy from the Fifth Dimension  }
\vskip .5cm

\centerline{   A. Stern\footnote{astern@ua.edu} and Chuang Xu\footnote{cxu24@crimson.ua.edu}}

\vskip 1cm

\begin{center}
  { Department of Physics, University of Alabama,\\ Tuscaloosa,
Alabama 35487, USA\\}

\end{center}

\vskip 0.5cm

\abstract{
After extending the Regge-Teitelboim formulation of gravity to include the case where the background embedding space is not flat, we examine the dynamics of the four-dimensional   $k=0$ Robertson-Walker (RW) manifold embedded in various five-dimensional backgrounds.   We find that when   the  background is five-dimensional de Sitter space, the RW manifold undergoes a transition from a de-accelerating phase to an accelerating phase.  This  occurs before the  inclusion of  matter, radiation or cosmological constant sources, and thus does not require a balance of  different components.  We obtain a reasonable two-parameter fit  of this model to the Hubble parameter data.}

\newpage

Regge-Teitelboim (RT) gravity is an alternative to standard general relativity
where the dynamical degrees of freedom are associated with the  embeddings of  the space-time manifold in a fixed higher dimensional background.\cite{Regge:2016gaw,Deser:1976qt,Pavsic:1984kv,Tapia:1988ze,Maia:1989du,Bandos:1996rw,Capovilla:2006dc}   Solutions to Einstein equations satisfy the RT field equations.  More generally, the RT  formulation of gravity can effectively produce source terms in the standard Einstein equations that are not attributable to the energy-momentum tensor, but rather  are a result of the embedding.   These RT source terms have the potential of providing  an explanation for certain cosmological phenomena, such as    for cosmic acceleration.  This idea was entertained in \cite{Fabi:2022slr}, where a toy model was presented.\footnote{See also \cite{Davidson:1997ys}
. }  There it was shown that RT gravity can generate  cosmic acceleration for a simple class of embeddings of the $k=-1$ Robertson-Walker (RW) metric in a flat five-dimensional background. Moreover, a transition from a de-accelerating to accelerating phase could be observed in a specific, but not  physical, example.  No cosmological acceleration was found for the case of the $k=1$ RW metric, and  the currently favored $k=0$ case was not considered. 

This article  attempts to apply RT gravity in the direction of a more realistic  model of  the observed cosmological acceleration. The approach taken here relies on embedding the RW manifold in a curved background.  While previous discussions of RT gravity have been restricted to  flat backgrounds,  the formalism can easily  be extended to curved background spaces, as is demonstrated here.  After developing the formalism, we then  apply it to cosmology by embedding the four-dimensional  RW manifold    in three different five-dimensional background spaces.  We specialize to $k=0$, although the other cases can also be considered as well.  The backgrounds we consider are: $i)$ $R^{4,1}$,  $ii)$  $AdS_5$   and $iii)$ $dS_5$.  As a first approximation, we obtain the evolution of the scale factor on the RW manifold in the absence of  matter, radiation or cosmological constant sources.  
We get that the acceleration of the scale factor is negative  for all time for cases $i)$ and $ii)$.  On the other hand,   for  case $iii)$ we find that a transition from the de-accelerating phase to an accelerating phase occurs at a finite time.   The evolution in this case 
 is determined by two free parameters,  the curvature of the background de Sitter space and the strength of the RT source term.  The two parameters allow for  a fit  to the  Hubble parameter data. Unlike in the  $\Lambda$CDM model, neither the matter density nor the cosmological constant play a role in the fit, meaning that their contributions  should be significantly weaker than the RT source term, and furthermore, that they can have arbitrary  strength relative to each other.  So here we are able to avoid the  coincidence puzzle of the $\Lambda$CDM model, where the  matter contribution to Einstein equations, coincidentally, is of the same order of magnitude as the cosmological constant contribution at the current time. 

We begin with a very brief discussion of RT gravity, or more precisely, its generalization to the case where the $d-$dimensional  background space ${\bf M}_d$,  $d>4$,  is  not necessarily flat.  We denote a local set of coordinates on ${\bf M}_d$ by $Y^a$,   $a,b,\cdots=0,...,d-1$, and its associated metric tensor metric by  ${\tt \bf g}_{ab}(Y)$.  Next embed a four-dimensional space-time  manifold ${\cal M}_4$  in ${\bf M}_d$.   This can be done by introducing the set of functions  $Y^a=Y^a(x)$, where $x^\mu$,   $\mu,\nu,\cdots=0,...,3$, span  ${\cal M}_4$.
The metric tensor $ g_{\mu\nu}(x)$ on ${\cal M}_4$  is defined to be induced from  ${\tt \bf g}_{ab}(Y)$. So 
\be  g_{\mu\nu}(x)={\tt \bf g}_{ab}(Y)\partial_\mu Y^a\partial_\nu Y^b\;,\label{induced}\ee
 $\partial_\mu$ denoting differentiation with respect to $x^\mu$.  As is usual  $ g_{\nu\lambda}$  is required to be invertible, and  metric compatible on ${\cal M}_4$, $\nabla _\mu g_{\nu\lambda}=0$,  $\nabla _\mu$ being the covariant derivative on ${\cal M}_4$.  The latter leads to  the identity:
\be {\tt \bf g}_{ab}\nabla_\lambda \partial_\mu Y^a\,\partial_\nu Y^b+\frac 12\frac{\partial{\tt \bf g}_{ab}}{\partial Y^c}\left(\partial_\mu Y^a \partial_\nu Y^b\partial_\lambda Y^c+\partial_\nu Y^a \partial_\lambda Y^b\partial_\mu Y^c-\partial_\lambda Y^a\partial_\mu Y^b \partial_\nu Y^c\right)=0\label{idntepto}\ee
To derive this compute $\nabla _\lambda g_{\mu\nu}+\nabla _\mu g_{\nu\lambda}-\nabla _\nu g_{\lambda\mu}$ using (\ref{induced}), and apply metric compatibility and the Leibniz rule.

 RT gravity  assumes the usual Einstein-Hilbert action $S_{EH}$ for the  gravitational field, however the dynamical degrees of freedom are the embedding functions, not $g_{\mu\nu}$.  So upon including source terms  $ S_{source}$, one has
 \be S=S_{EH}+ S_{source}, \qquad S_{EH}=\frac 1{16\pi G}\int_M d^4x\sqrt{|g|} R\;\label{actnwthsrs}\;,\ee
with the scalar curvature constructed from (\ref{induced}).  Field dynamics is obtained from variations of $Y^a$.  This gives 
$$ \partial_\mu\left(\sqrt{|g|}E^{\mu\nu}{\tt \bf g}_{ab}\partial_\nu Y^b\right)-\frac 12\sqrt{|g|} E^{\mu\nu}\frac{\partial {\tt \bf g}_{bc}}{\partial Y^a} \partial_\mu Y^b\partial_\nu Y^c=0\;, $$\be\qquad\quad E^{\mu\nu}=G^{\mu\nu}-8\pi G\,T^{\mu\nu}\;,\label{RTeq}\ee
$G^{\mu\nu}$  and $T^{\mu\nu}$ being the Einstein tensor and stress-energy tensor, respectively. As in Einstein gravity,
  $T^{\mu\nu}$ must be covariantly conserved.  To see this one can first re-write the field equations as
\be \nabla_\mu(E^{\mu\nu}{\tt \bf g}_{ab}\partial_\nu Y^b)-\frac 12 E^{\mu\nu}\frac{\partial {\tt \bf g}_{bc}}{\partial Y^a} \partial_\mu Y^b\partial_\nu Y^c=0\;,\ee
and then expand the first term using the Bianchi identity to obtain
\be -8\pi G\,\nabla_\mu T^{\mu\nu}\,{\tt \bf g}_{ab}\partial_\nu Y^b+E^{\mu\nu}\left(\nabla_\mu({\tt \bf g}_{ab}\partial_\nu Y^b)-\frac 12 \frac{\partial {\tt \bf g}_{bc}}{\partial Y^a} \partial_\mu Y^b\partial_\nu Y^c\right)=0\ee
Finally contract with $\partial_\lambda Y^a$
and apply (\ref{idntepto}) to get  $\nabla_\mu T^{\mu}_{\;\;\,\lambda}=0$.
The field equations (\ref{RTeq}) are obviously satisfied for  solutions to Einstein equations, $E^{\mu\nu}=0$.  More generally, $E^{\mu\nu}$ need not vanish.  Alternatively, we can argue that  the Einstein equations effectively pick up  additional source terms, which we denote by $T_{{}_{\rm RT}}^{\mu\nu}$,   which are not associated with the standard stress-energy tensor but rather are due to the embedding in the background space,
\be G^{\mu\nu}=8\pi G\,(T^{\mu\nu}+T_{{}_{\rm RT}}^{\mu\nu})\;,\label{mdfdeeq}\ee
Obviously, $T_{{}_{\rm RT}}^{\mu\nu}$ is covariantly conserved since  $T^{\mu\nu}$ is.

Next we  want to apply this dynamical system to the case where the embedded manifold ${\cal M}_4$ is that of standard cosmology, i.e., it is given by the  RW metric tensor.  Here we will specialize to the currently favored case of $k=0$
\be ds^2=-dt^2+ a(t)^2 dx^idx^i\label{k0metric}\;,\ee
where $t=x^0$ and $a(t)$ is the scale factor.  As a first approximation let us consider  source free RT gravity, i.e.,  $T^{\mu\nu}=0$.  From (\ref{mdfdeeq}) we know that the Einstein tensor need not vanish.  $T_{{}_{\rm RT}}^{\mu\nu}$ in (\ref{mdfdeeq})  needs to be computed from the particular choice of embedding, however from consistency with homogeneity and isotropy, we anticipate that its form should be analogous to that of a perfect fluid in the co-moving frame
\be T^{00}_{{}_{\rm RT}}=\rho_{{}_{\rm RT}}\qquad T^{11}_{{}_{\rm RT}}=T^{22}_{{}_{\rm RT}}=T^{33}_{{}_{\rm RT}}=a(t)^2 p_{{}_{\rm RT}}\;,\label{TndTRT}\ee
with $\rho_{{}_{\rm RT}}$ and $ p_{{}_{\rm RT}}$ being functions of $t$.
Since it  is covariantly conserved we have
\beqa 
 \dot \rho_{{}_{\rm RT}}+3\frac{\dot a}a(\rho_{{}_{\rm RT}}+p_{{}_{\rm RT}})&=&0\label{cntnuitRT}\;,
\eeqa  
the dot denoting a $t-$derivative.

Substituting(\ref{k0metric}) into (\ref{RTeq}) gives
\be \partial_t\left(F_1(t)  {\tt \bf g}_{ab}\partial_t Y^b\right)-\frac 12F_1(t) \partial_t Y^b\partial_t Y^c\frac{\partial {\tt \bf g}_{bc}}{\partial Y^a}=F_2(t)\left(\partial_i\left({\tt \bf g}_{ab}\partial_i Y^b\right)-\frac 12 \partial_i Y^b\partial_i Y^c\frac{\partial {\tt \bf g}_{bc}}{\partial Y^a}\right)\;,\quad\label{ptsx}
\ee
where \be   F_1(t)= 3a \dot a^2\qquad\quad F_2(t)=2\ddot a+\frac{\dot a^2}a
\label{F1ndF2}\ee
  (\ref{ptsx}) can produce equations for $\dot a$ and $\ddot a$ which can be written in the form of  $k=0$ Friedmann equations
\beqa   \frac {\dot a^2}{a^2}&=&\frac{ 8\pi G}3 \rho_{\tt RT}\label{Feq1}\\&&\cr
 \frac{\ddot a}a&=&-\frac{4\pi G}3(\rho_{{}_{\rm RT}}+3 p_{{}_{\rm RT}})\;,\eeqa  allowing us to identify $ \rho_{\tt RT}$ and $ p_{{}_{\rm RT}} $ in (\ref{TndTRT}).  The resulting expressions for  $ \rho_{\tt RT}$ and $ p_{{}_{\rm RT}} $ will in general depend on the background space and  the choice of embedding, as we illustrate in the examples that follow. 
 As stated previously, the  background spaces we consider are $R^{4,1}$,  $AdS_5$ and  $dS_5$.  We  use the same expression for the embedding in all three cases:
\be  \begin{pmatrix}
Y^0 \\Y^1 \\Y^2\\Y^3\\Y^4
\end{pmatrix}
=\begin{pmatrix}b(t)\\x^1\\x^2 \\x^3\\h(t)\end{pmatrix}\;,\label{mbdng2}\ee
where the functions $b(t)$ and $h(t)$ need to satisfy certain  constraints in order to recover the $k=0$ Robertson-Walker metric on the embedded four-dimensional manifold.

We next deduce  $ \rho_{\tt RT}$ and $ p_{{}_{\rm RT}} $ for the three different cases.
\begin{enumerate}

\item Flat $5-$dimensional background $R^{4,1}$

A trivial system results if one chooses Cartesian coordinates for $R^{4,1}$ and maps  to ${\cal M}_4$ using (\ref{mbdng2}), as this restricts the scale factor in (\ref{k0metric}) to be one.
Alternatively, a nontrivial function $a(t)$ can result from a different coordinatization on $R^{4,1}$ , such as is  in \cite{Robinson:1987mr}, \cite{Akbar:2017vja} where 
\be (ds^2)_{R^{4,1}}=-(dY^0)^2+(Y^0+ Y^4)^2\left( (dY^1)^2+(dY^2)^2+(dY^3)^2\right)+(dY^4)^2\ee 
It can be checked that the five-dimensional curvature resulting from this metric is zero.  Now  using (\ref{mbdng2}) to map to (\ref{k0metric}) one gets that $b(t)$ and $h(t)$ should satisfy
\be   b(t)+h(t)=a(t)\qquad\quad \dot b^2- \dot h^2=1\;\label{andb2h2f2}\ee 
Now substituting (\ref{mbdng2})  in (\ref{ptsx}) gives
\beqa  \partial_t(\dot b F_1)&=&\;\;\,3 F_2 a\\ \partial_t(\dot h F_1)&=&-3 F_2 a\label{pt16}\eeqa  The sum of these two equations leads to a constant of motion $\partial_t(\dot a F_1)=0$,
from which we get the following expression for $\rho_{\tt RT}$
\be   \rho_{\tt RT}=\frac{c_0}{a^3\dot a}\;,\label{mdfidF1}\ee
$c_0$ being a constant.   The Friedmann equation (\ref{Feq1}) then gives $\dot a^3\propto \frac 1 a$, and so there is  no acceleration as $a$ increases.  One gets a simple solution for the scale factor in this case:  $a(t)\propto  t^{3/4}$ for $a(0)=0$.  This coincides with the time evolution of the scale factor in the presence of a perfect fluid with equation of state $p=-\frac 19 \rho$.  The same result was observed in \cite{Davidson:1997ys} for a different choice of embedding.

\item $AdS_5$  background

Here we cover a patch of  $AdS_5$ using Poincar\'e coordinates.  The background metric is
\be (ds^2)_{AdS_5}=-\frac{(Y^4)^2 }{L^2}(dY^0)^2+\frac{ (Y^4)^2}{L^2}\Bigl( (dY^1)^2+ (dY^2)^2+ (dY^3)^2\Bigr)+\frac{L^2(dY^4)^2}{(Y^4)^2} \;, \ee
the constant $L$ denoting the $AdS_5$ radius of curvature. Utilizing the embedding (\ref{mbdng2}), the $k=0$ RW metric (\ref{k0metric}) is recovered provided  that 
\be h=L \,a\qquad\quad a^2\dot b ^2-L^2\frac{\dot a^2}{a^2}=1\ee
Substituting (\ref{mbdng2})  in (\ref{ptsx}) gives 
\beqa  \partial_t(a^2\dot b F_1)&=&0\label{k51eq1}\\ &&\cr
L^2\partial_t\left(\frac {\dot a}{a^2} F_1\right)+\left(a\dot b^2+L^2\frac{\dot a^2}{a^3}\right)F_1&=&-3 a F_2 \label{k51eq2}\eeqa
From  (\ref{Feq1}) and  (\ref{k51eq1}) we then get
\be  \rho_{\tt RT}=\frac{c_0}{a^3\sqrt{L^2\dot a^2+a^2}}\;,\label{mdfidF2}\ee
Note that the form  (\ref{mdfidF1}) resulting from the flat background is recovered in the limit $L\rightarrow\infty$, or more precisely when $|\frac{\dot a}a|>>\frac 1L$. 

\item $dS_5$  background

Using the so-called flat slicing the metric  for $dS_5$   is
\be (ds^2)_{dS_5}=-(dY^0)^2+e^{2Y^0/L}\Bigl( (dY^1)^2+ (dY^2)^2+ (dY^3)^2\Bigr)+e^{2Y^0/L} (dY^4)^2\;,\ee  $L$ again being the radius of curvature.
Now (\ref{k0metric}) is recovered from the embedding (\ref{mbdng2}) for
\be e^{b/L}=a\qquad\quad  L^2\frac{\dot a^2}{a^2}-a^2\dot h^2=1\ee
After substituting (\ref{mbdng2})  in (\ref{ptsx}) 
\beqa 
L^2\partial_t\left( \frac{\dot a}a F_1\right)+a^2\dot h^2 F_1-3 a^2 F_2 &=&0\\  \partial_t(a^2\dot h F_1)&=&0\label{dS22ndeq}\eeqa 
From  (\ref{Feq1}) and  (\ref{dS22ndeq}) we then get
\be   \rho_{\tt RT}=\frac{c_0}{a^3\sqrt{L^2{\dot a^2}-{a^2}}}\;\label{mdfidF1}\ee
$c_0$  is real which means we need that $|\frac{\dot a}a|>\frac 1L$.  The expression  (\ref{mdfidF1})  is once again recovered  for $|\frac{\dot a}a|>>\frac 1L$.

\end{enumerate}

\medskip

\noindent
To summarize, the source term  $\rho_{\tt RT}$ for the three  different backgrounds has the form\footnote{ Here we have done  a rescaling of the constant $c_0$ for the case $k_5=0$.}
\be  \rho_{\tt RT}=\frac{c_0}{a^3\sqrt{L^2{\dot a^2}-k_5{a^2}}}\;,\label{rhort}\ee
where $k_5$  defines the curvature of the five-dimensional background space: $k_5=0,-1,1$ for  $R^{4,1}$, $AdS_5$ and  $dS_5$, respectively.  Moreover, from (\ref{Feq1}) one has  that \be  {\dot a^2}a\sqrt{L^2{\dot a^2}-k_5{a^2}}={\rm constant}\label{cnstfmtn}\ee 
$p_{\tt RT}$ can be determined from the conservation  law (\ref{cntnuitRT}) leading to a time-dependent\footnote{ The case $k_5=0$ is an exception.  After using (\ref{cnstfmtn}) one gets the simple relation $p_{{}_{\rm RT}}=-\frac 19 \rho_{\tt RT}$.} equation of state
\be p_{{}_{\rm RT}}=-\frac a{3\dot a}\dot\rho_{{}_{\rm RT}}-\rho_{{}_{\rm RT}}=\frac{a \left(L^2\ddot a-k_5 a\right)}{3 \left(L^2\dot a^2-k_5
   a^2\right)}\rho_{\tt RT}\label{pressurert}\ee

The   evolution of the scale factor for the three cases $k_5=1,0,-1$ is obtained   from (\ref{cnstfmtn}).  As stated previously, for  $k_5=0$ one  gets $a(t)\propto  t^{3/4}$.  We resort to numerical integration to obtain solutions  for the other two cases,  $k_5=\pm 1$. The results for all  three cases 
are plotted in Fig. 1, using the initial condition $a(0)=0$. All three cases agree for small $t$, i.e. $a(t)\propto  t^{3/4}$ as  $L\frac{\dot a }a \rightarrow \infty$, and so $\ddot a<0$. For cases  $k_5=0$ and $-1$, we find that $\ddot a<0$,  {\it for all }$t$.   The situation is more interesting for  $k_5=1$, corresponding to the de Sitter background.  In this case,   $\ddot a$ vanishes  at finite $t$, when $L\frac{\dot a }a=\sqrt{2} $, thus signaling a transition from the de-accelerating phase to an accelerating phase. We get that $L\frac{\dot a }a$ goes asymptotically to one in the $t\rightarrow\infty$  limit, where the scale factor undergoes an exponential expansion at leading order, 
\be a(t)\rightarrow a_1 e^{t/L}\left( 1-a_2 e^{-8t/L}+\cdots\right)\;,\qquad {\rm as}\;t\rightarrow\infty\;,\ee
 $a_1$ and $a_2$ being positive constants.  From (\ref{pressurert}) we can obtain the equation of state for the RT source as a function of time.  The ratio
$p_{{}_{\rm RT}}/\rho_{{}_{\rm RT}}$, standardly denoted by $w$, goes from $-\frac 19$, near $t=0$, to $-\frac 13$, at the transition, to $-1$, in the limit $t\rightarrow \infty$.
Note that unlike in the $\Lambda$CDM model, here we get a transition from the de-accelerating phase to an accelerating phase even without  the inclusion of a matter component or cosmological constant component to the Friedmann equations.
  \begin{figure}[!h]
\begin{center}
\includegraphics[height=2.5in,width=2.5in,angle=0]{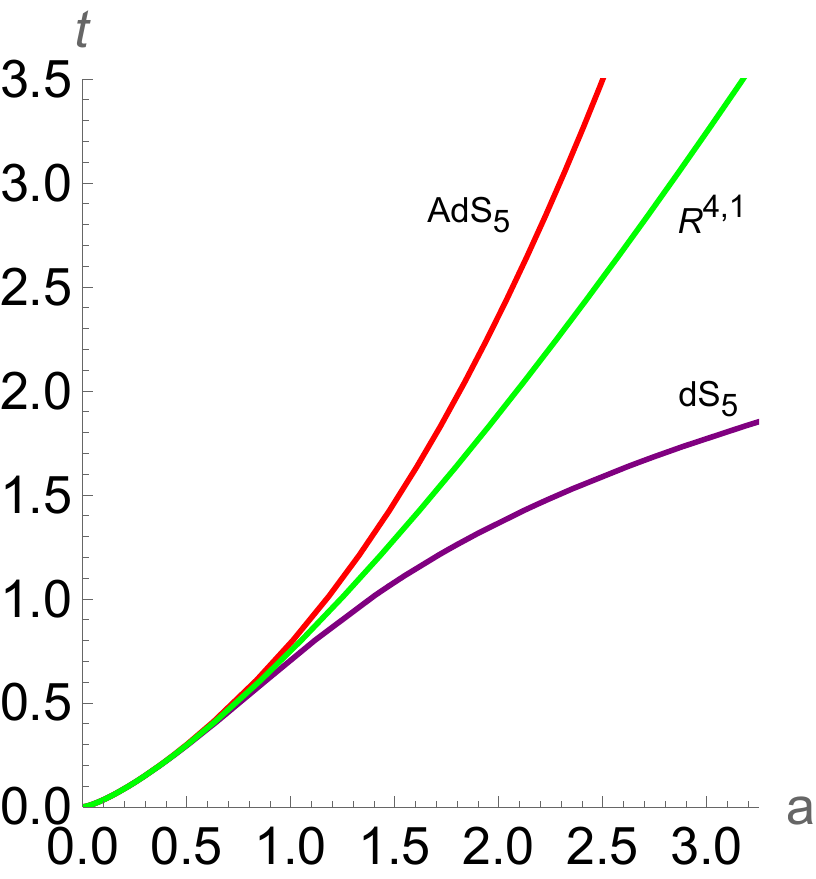}
\caption {Plot of $t$ vs $a$ for three different five-dimensional background spaces: $R^{4,1}$, $AdS_5$ and  $dS_5$. (Here we set $L=1$.)}
\end{center}
\end{figure}

\begin{figure}[!h]
\centering     
\subfigure[$H$ versus $z$]{\label{fig:a}\includegraphics[height=71mm,width=68mm]{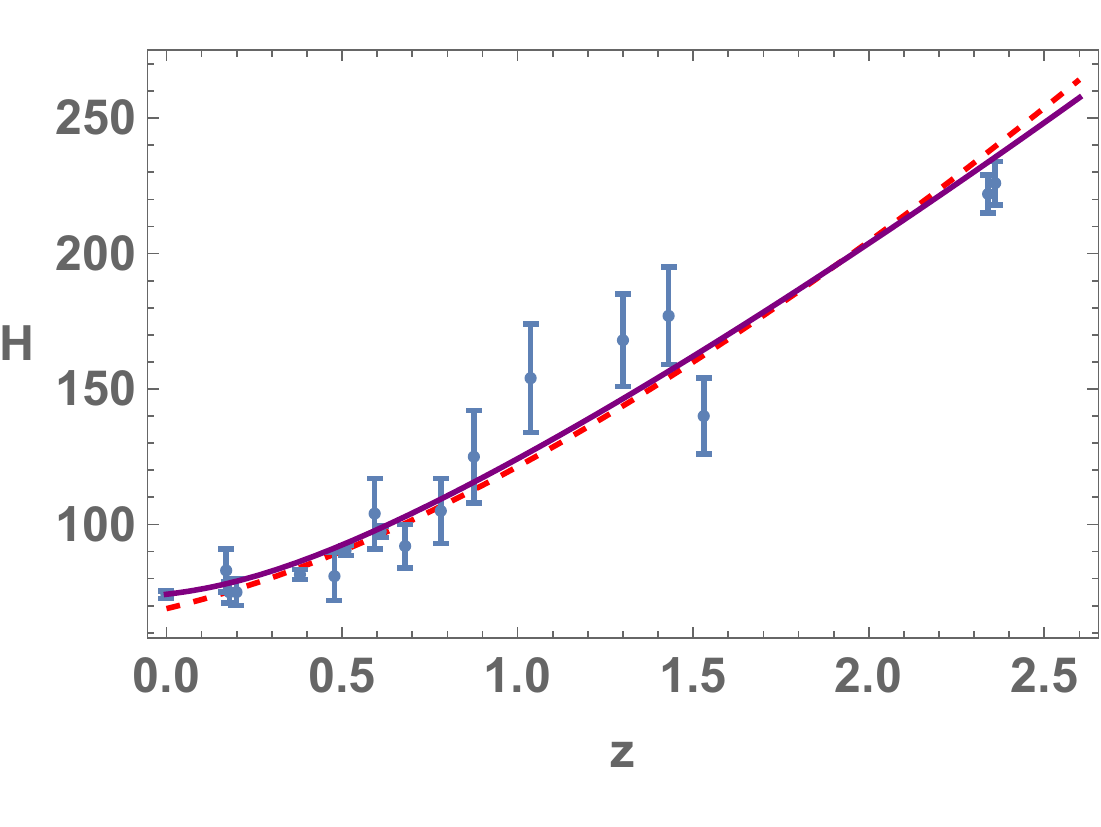}}
\hspace{10.00mm}
\subfigure[$ H/(1+z)$ versus $z$]{\label{fig:b}\includegraphics[height=74mm,width=70mm]{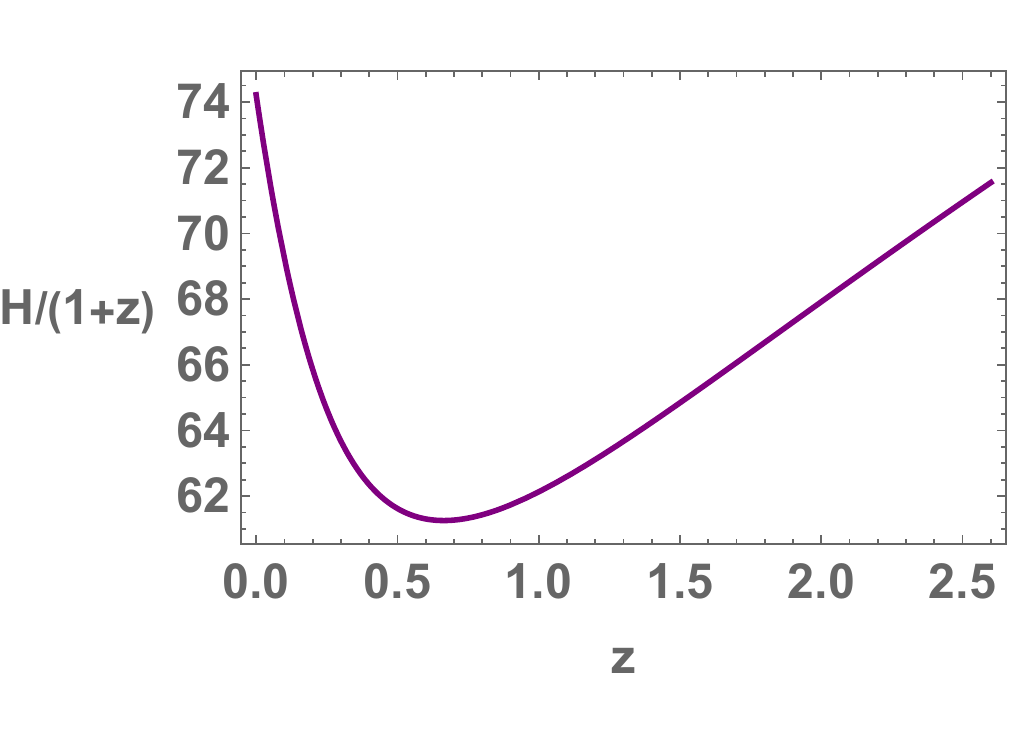}}
\caption{ The solid purple curve  in figure (a) represents a fit of  eq. (\ref{Hvszeq}) with the Hubble parameter data, while the dashed red curve is $\Lambda$CDM. $H$ is given in units of $ {\rm km\, s}^{-1}{\rm Mpc}^{-1}$. The best fit occurs for $\tilde c_0\approx .26$ and $1/L\approx 72\,{\rm km\, s}^{-1}{\rm Mpc}^{-1}$.   From figure (b) the minimum of $H/(1+z)$ for the  best fit occurs at $z\approx .675$, corresponding to the transition from a de-accelerating phase to an acceleration phase.}
\end{figure}

Finally, we proceed with a  fit of the $k_5=1$ case to observational data.  (\ref{cnstfmtn})
 gives an algebraic  relation between the Hubble parameter $H=\dot a/a $ and the redshift parameter $z=a_0/a-1$, where $a_0$ is the scale parameter at the current time.  It is
\be  L^2 H^2{\sqrt{ L^2 H^2-1}}={\tilde c_0\, (1+z)^4}\;,\label{Hvszeq}\ee
where  $\tilde c_0=\frac{ 8\pi G}3 L^2 a_0^{-4} c_0$.  In Fig. 2(a) we fit the real solution to eq. (\ref{Hvszeq}) to  observed results for $H$ versus $z$ using the data in Table 1.  The best fit occurs for  $\tilde c_0\approx .26$ and $1/L\approx 72 \,  {\rm km\, s}^{-1}{\rm Mpc}^{-1}$.  For $H$ evaluated at $z=0$ one gets $H(0)\approx 74  \, {\rm km\, s}^{-1}{\rm Mpc}^{-1}$.  Our fit in Fig. 2(a) is compared to that of $\Lambda$CDM, where the expression for the Hubble parameter is given by $H=H_0 \sqrt{(1+z)^3\Omega_m +\Omega_\Lambda}$, with $\Omega_m=.3$,  $\Omega_\Lambda=.7$  and $H_0\approx 68.92 \, {\rm km\, s}^{-1}{\rm Mpc}^{-1}$.
 $H/(1+z)$ (which is proportional to $\dot a$) versus $z$ is plotted in Fig. 2(b), using our fit for $H$ in Fig 2(a). It shows that the transition from a de-acceleration phase to acceleration phase occurs at $z\approx .675$, which is similar to the value predicted by $\Lambda$CDM.

We now summarize some of the features of this model.    After generalizing RT gravity to curved backgrounds, we found universal formulas for the effective density and pressure,  (\ref{rhort}) and (\ref{pressurert}), respectively, resulting from embedding  the $k=0$ RW manifold in three different five-dimensional background spaces.  We suspect that the results found here are dependent on the choice of embedding (in addition to the choice of background space), although we have not found specific examples of this.

A reasonable fit to the Hubble parameter data was obtained in the case where the background was  de Sitter space.  This is true even without considering the usual stress-energy contributions to the Einstein equations, which on the other hand, play an essential role for $\Lambda$CDM.     Such components can  easily be included in our model by adding   appropriate terms  to (\ref{F1ndF2}) and consequent equations.  For the case of nonrelativistic matter, one ends up with the following modification to (\ref{Hvszeq}):
\be \frac{L^2H^2 }{(1+z)^3}-\frac{\tilde c_0 (1+z)}{\sqrt{L^2H^2- 1}}={\tilde c_1}\;,\label{Hubwthmtr}\ee
where  $\tilde c_1$ is an additional constant which quantifies the nonrelativistic matter component.   The inclusion of the additional parameter  $\tilde c_1$ does not appear  to improve the previous fit in any  significant manner.

The presence of the square root in (\ref{Hvszeq}) [and also in (\ref{Hubwthmtr})] gives a lower bound on the Hubble parameter, $H(z)>1/L$, which is in agreement with observation.

The fit we obtained to the Hubble parameter data holds for values of  $z$ up to approximately $2.36$.  Concerning $z>2.36$, the deviation of our fit in Fig. 2  with that of $\Lambda$CDM grows when extrapolating to higher $z$.  However, our fit did not include  contributions from the stress-energy tensor, which can play a more significant role at large $z$. For example, if one considers the matter density $\rho_m$ which is proportional to $ a^{-3}$, then its relative contribution is  $\rho_m/\rho_{\tt RT}\propto \sqrt{L^2 H^2-1}/(z+1)$, which grows like $LH/z$ for large $z$. Also, there is no reason to assume  that the $5$d de Sitter background is valid for all $z$.  This among other issues is open for further investigation/speculation.

\begin{table}[!h]
\centering
\footnotesize
\begin{tabular}{|l|l|l|l|l|l|l|}

\hline
$z$&$H$&$\sigma_H$&\\
\hline
0   &$74.03$&$1.42$  &  \cite{Riess:2019cxk}   \\

\hline
.17&$83$&$ 8$&\cite{DStern} \\
\hline
.1791&$ 75$&$4$& \cite{Moresco:2012jh} \\
\hline
.1993&$75$&$5$& \cite{Moresco:2012jh} \\

\hline
.38&$ 81.5$ &$1.9 $&\cite{BOSS:2016wmc}  \\

\hline
.4783&$ 80.9$&$ 9$&\cite{Moresco:2016mzx} \\
\hline
.51 & $90.4$& $ 1.9$&\cite{BOSS:2016wmc}  \\

\hline
.5929&$104$&  $13$ &\cite{Moresco:2012jh}  \\

\hline
  .61&$97.3$&  $2.1$ &\cite{BOSS:2016wmc}    \\
\hline
  .6797&$92$&  $8$ &\cite{Moresco:2012jh}     \\
\hline

 .7812&$105$&  $12$ &\cite{Moresco:2012jh}     \\
\hline
 .8754&$125$&  $17$ &\cite{Moresco:2012jh}     \\
\hline
 1.037&$154$&  $20$ &\cite{Moresco:2012jh}     \\
\hline
 1.3&$168$&  $17$ &\cite{DStern}     \\
\hline
 1.43&$177$&  $18$ &\cite{DStern}     \\
\hline
 1.53&$140$&  $14$ &\cite{DStern}     \\
\hline
 2.34&$222$&  $7$ &\cite{BOSS:2014hwf}    \\
\hline
 2.36&$226$&  $8$ &\cite{BOSS:2013igd}   \\
\hline

\end{tabular}
\caption{Data used for fit in Fig. 2.  Columns 1-4 are $z$, $H$ , error in $H$ and citation respectively. Columns 2\&3 are in units of  ${\rm km\, s}^{-1}{\rm Mpc}^{-1}$. Data was selected with $\sigma_H<.15 H$.}

\label{tab:template}

\end{table}

\end{document}